\documentclass{article}
\usepackage[margin=1in]{geometry}
\usepackage{amsmath}
\usepackage{graphicx}
\usepackage{amsfonts}
\usepackage{hyperref}
\DeclareMathOperator{\si}{si}
\DeclareMathOperator{\ci}{ci}
\DeclareMathOperator{\ti}{ti}
\title{A Simple Model for Quantum Gravity I:\\the one-dimensional case}
\author{Ricardo Paszko\footnote{ricardo.paszko@ufabc.edu.br}\\
Universidade Federal do ABC}
\begin{document}
\maketitle
\begin{abstract}
We propose and analytically solve a simple Euclidean model for quantum gravity in one dimension. The geometry is parametrized by angles associated with a circle of fixed radius $\ell$, allowing the use of the normalized Haar measure of U(1) and character expansions. In the case of an open curve, the continuum limit is trivial: the size of the universe is infinite, independently of the value of the cosmological constant. In the closed case, however, the continuum limit is nontrivial, resulting in a finite-size universe for positive cosmological constant. The presence of a scalar field slightly modifies this result while preserving its behavior. We also discuss possible generalizations of the model to higher dimensions.

\end{abstract}
\section{Introduction}
Consider the simplest spacetime: a spacetime with only time and no space. Assume that this ``time'' is positive definite, that is, Euclidean time. In the closed case, let us also consider it periodic. Therefore, this spacetime can be visualized as a continuous simple closed curve embedded in a two-dimensional space.

An approximation of this curve by $N$ straight line segments, that is, a one-dimensional simplicial approximation, was studied in the 1990s using \emph{Regge calculus} \cite{Hamber} with numerical calculations, performed by computers. Choosing the length $l_n$, for $n=1,2,3,\ldots,N$, of these line segments, varying from $\epsilon\leq l_n<\infty$, for some small length $\epsilon$. In addition, an integration measure $\prod_{i=1}^Ndl_n^2l_n^\sigma$ is chosen for some power $\sigma$.

However, analytical expressions can be obtained only in the simplest case of an open curve, which has a trivial continuum limit (compare Eq.~(\ref{trivial}) below with the corresponding result in Ref.~\cite{Hamber}, taking $\epsilon\to0$ and $\sigma>-2$). In the closed case, even the simplest configuration, $N=3$, is subject to the triangle inequalities
\[l_1<l_2+l_3,\qquad l_2<l_3+l_1,\qquad\text{and}\qquad l_3<l_1+l_2,\]
which prevent an analytical calculation for general $\epsilon$ and $\sigma$. For $N\geq4$, the inequalities become increasingly complicated. Since the continuum limit requires $N\to\infty$, these difficulties are further amplified, and in higher dimensions they make it impossible to obtain analytical expressions within this formulation.

In the present article, we introduce a simple model to avoid numerical calculations, problems with inequalities, and ambiguities in the values of $\epsilon$ and $\sigma$ within Regge calculus. Instead of integrating over the edge lengths directly, we parametrize them by angles associated with a circle of fixed radius $\ell$. This allows us to obtain analytical expressions even in the closed case and for general $N$ in one dimension. Our model can also be generalized to spacetimes of higher dimensions, as we will briefly discuss.

\section{A Simple Model}
\subsection{Pure Gravity}
\begin{figure}
\centering
\includegraphics[scale=0.2]{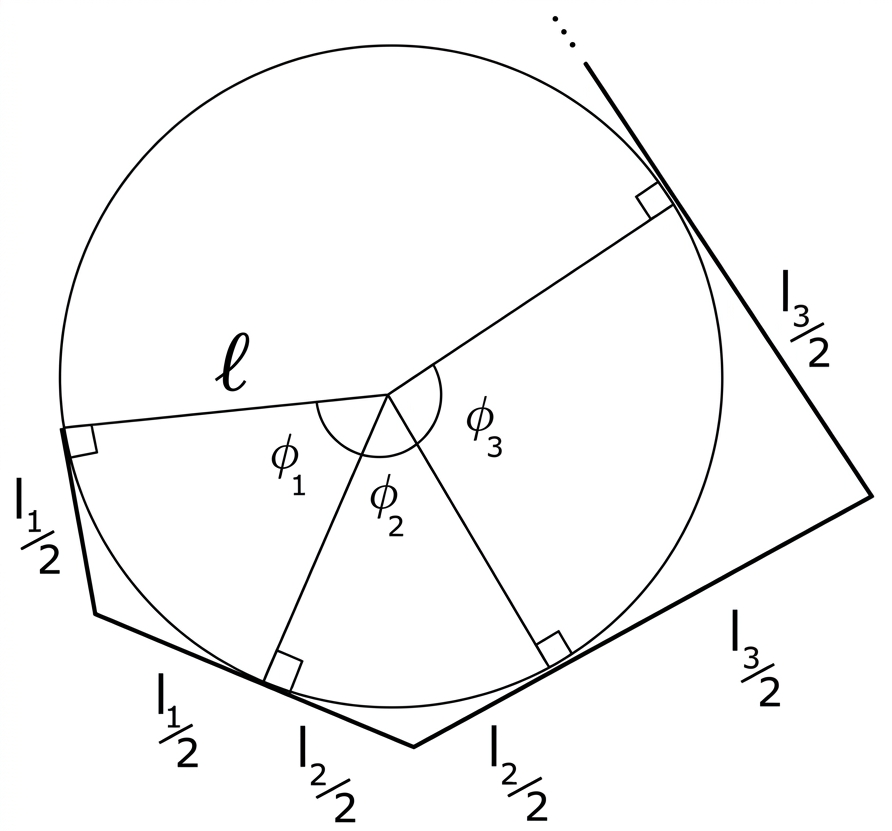}
\caption{A curve constructed around a circle of radius $\ell$ using straight lines.}
\label{curve}
\end{figure}

Imagine we construct the straight lines tangent to a circle of radius $\ell$ as in Fig.~(\ref{curve}).
Notice that $\ell$ is the only constant in our model. This is similar in spirit to the equilateral triangle of edge length $a=\text{constant}$ in the numerical calculations of the so-called \emph{causal dynamical triangulations} \cite{Loll}.

Here, all lengths $l_n=2\ell\tan(\phi_n/2)$ can vary from $0\leq l_n<\infty$, as the angles vary from $0\leq\phi_n\leq\pi$. In the closed case we have a constraint
\begin{equation}
\sum_{n=1}^N\phi_n=2\pi.\label{constraint}
\end{equation}
Since the Ricci scalar vanishes identically in 1D, the only interesting term that can be put in the gravitational action is the total length of the curve
\begin{equation}
L=\sum_{n=1}^Nl_n.
\end{equation}

Thus, a natural partition function $Z$ for pure gravity, in the open case, is
\begin{align}
Z&=\left(\prod_{n=1}^N\int_0^\pi\frac{d\phi_n}{\pi}\right)\exp(-\lambda L)\nonumber\\
&=\left[\int_0^\pi\frac{d\phi}{\pi}\exp\left(-2\lambda\ell\tan\frac{\phi}{2}\right)\right]^N\nonumber\\
&=\left\{\frac{2}{\pi}[\ci(2\lambda\ell)\sin(2\lambda\ell)-\si(2\lambda\ell)\cos(2\lambda\ell)]\right\}^N,
\label{open}
\end{align}
where $\lambda$ is the cosmological constant and the result is very simple in the open case since the integrals factor out.

Observe that the expectation value of the total length is
\begin{align}
\langle L\rangle=&\frac{1}{Z}\left(\prod_{n=1}^N\int_0^\pi\frac{d\phi_n}{\pi}\right)L\exp(-\lambda L)\nonumber\\
=&-\frac{d(\ln Z)}{d\lambda}\label{expectation}
\end{align}
and the continuum limit is trivial: $\langle L\rangle$ goes to infinity,
\begin{equation}
\langle L\rangle=2N\ell\left[\frac{1+\ti(2\lambda\ell)\tan(2\lambda\ell)}{\ti(2\lambda\ell)-\tan(2\lambda\ell)}\right]\to\infty,\label{trivial}
\end{equation}
as $N\to\infty$. Here $\si$ and $\ci$ are the sine and cosine integrals, and we defined $\ti(x)=\si(x)/\ci(x)$.

Now, in the closed case, we have the constraint, Eq.~(\ref{constraint}), and thus
\begin{equation}
Z={\cal N}\left(\prod_{n=1}^N\int_0^\pi\frac{d\phi_n}{\pi}\right)\delta\left(\sum_{n=1}^N\phi_n-2\pi\right)\exp\left(-2\lambda\ell\sum_{n=1}^N\tan\frac{\phi_n}{2}\right),
\label{Z}
\end{equation}
where ${\cal N}$ is a normalization constant.

Also notice that, using absolute values in Eq.~(\ref{Z}), we can double the integration limits in $\phi_n$ to make use of the orthogonality relation
\begin{equation}
\int_{-\pi}^\pi\frac{d\phi}{2\pi}\chi_p(\phi)\chi_q^*(\phi)=\delta_{pq}
\label{orthogonality}
\end{equation}
between the U(1) characters $\chi_p(\phi)=\exp(-ip\phi)$ for $p\in\mathbb{Z}$.

Therefore, we can expand the delta function containing absolute values by employing a product of characters\footnote{The familiar expansion $\delta(\phi)=\sum_p\chi_p(\phi)/2\pi$ does not apply directly to multiple variables. For example, try to calculate $\int_{-\pi}^\pi d\phi_1\int_{-\pi}^\pi d\phi_2(\phi_1+\phi_2)\delta(\phi_1+\phi_2-\pi)$, which is equal to $\pi^2$, using the \emph{incorrect} expansion $\delta(\phi_1+\phi_2-\pi)\neq\sum_p\chi_p(\phi_1+\phi_2-\pi)/2\pi$, you will get zero.\label{Dirac}}
\begin{equation}
\delta\left(\sum_n|\phi_n|-2\pi\right)=\sum_{p_1,p_2,\ldots,p_N}c_{p_1p_2\ldots p_N}\chi_{p_1}(\phi_1)\chi_{p_2}(\phi_2)\ldots\chi_{p_N}(\phi_N),
\label{delta}
\end{equation}
where the coefficients $c_{p_1p_2\ldots p_N}$ can be found, using Eq.~(\ref{orthogonality}), to be
\begin{equation}
c_{p_1p_2\ldots p_N}=\int_{-\pi}^\pi\frac{d\phi_1}{2\pi}\int_{-\pi}^\pi\frac{d\phi_2}{2\pi}\ldots\int_{-\pi}^\pi\frac{d\phi_N}{2\pi}\delta\left(\sum_n|\phi_n|-2\pi\right)\chi_{p_1}^*(\phi_1)\chi_{p_2}^*(\phi_2)\ldots\chi_{p_N}^*(\phi_N).
\label{c}
\end{equation}

For example, with $N=3$, we have
\begin{equation}
c_{p_1p_2p_3}=
\begin{cases}
1/2\pi,& p_1=p_2=p_3=0\\
1/8\pi,& p_1=p_2=p_3\neq 0\\
[(-1)^{p_1+p_2}-1]p_1^2/\pi^3(p_1^2-p_2^2)^2,& p_1\neq p_2=p_3\\
[p_1^2(p_2^2-p_3^2)(-1)^{p_2+p_3}+\ldots]/\pi^3(p_1^2-p_2^2)(p_2^2-p_3^2)(p_3^2-p_1^2),& p_1\neq p_2\neq p_3
\end{cases}
\label{c3}
\end{equation}
where the ellipsis means cyclic permutations of indices $1,2,3$, and $c_{-p_1,p_2,p_3}=c_{p_1,p_2,p_3}$, etc. Similarly for $N\geq 4$.

Using Eq.~(\ref{orthogonality}) and an inverse Mellin transform
\[\exp(-x)=\int_{r-i\infty}^{r+i\infty}\frac{dz}{2\pi i}\frac{\Gamma(z)}{x^z},\]
where $r=\Re(z)>0$; see Ref.~\cite{Iwata}, for the exponential in Eq.~(\ref{Z}), we can expand the exponential as
\begin{equation}
\exp[-2\lambda\ell\tan(|\phi|/2)]=\sum_{p=-\infty}^\infty d_p\chi_p(\phi)
\label{exponential}
\end{equation}
with coefficients ($d_{-p}=d_p$)
\begin{align}
d_p=&\int_{-\pi}^\pi\frac{d\phi}{2\pi}\exp[-2\lambda\ell\tan(|\phi|/2)]\chi^*_p(\phi)\nonumber\\
=&\int_{r-i\infty}^{r+i\infty}\frac{dz}{2\pi i}\frac{\Gamma(z)}{(2\lambda\ell)^z}\int_{-\pi}^\pi\frac{d\phi\,\chi^*_p(\phi)}{2\pi\tan^z(|\phi|/2)}\nonumber\\
=&\int_{r-i\infty}^{r+i\infty}\frac{dz}{2\pi i}\frac{\Gamma(z)}{(2\lambda\ell)^z}\sec\left(\frac{\pi z}{2}\right)f_p(z),
\end{align}
where
\begin{equation}
f_p(z)=
\begin{cases}
1,&p=0\\
\frac{2^{2p-1}}{p!}\left(\frac{z+1}{2}\right)_p{}_3F_2\left(\frac{1}{2}-p,-p,-p;1-2 p,\frac{1}{2}-p-\frac{z}{2};1\right),&p\geq 1
\end{cases}
\end{equation}
in which $F$ is a hypergeometric function, and $(z)_p$ is a Pochhammer symbol.

For example,
\[d_0=\frac{2}{\pi}[\ci(2\lambda\ell)\sin(2\lambda\ell)-\si(2\lambda\ell)\cos(2\lambda\ell),\quad d_1=-\frac{4\lambda\ell}{\pi}[\ci(2\lambda\ell)\cos(2\lambda\ell)+\si(2\lambda\ell)\sin(2\lambda\ell)],\]
and so on\textellipsis This can be summarized as
\begin{align}
d_p=&\frac{2}{\pi}[\ci(2\lambda\ell)\sin(2\lambda\ell)-\si(2\lambda\ell)\cos(2\lambda\ell)]f_p(z)\nonumber\\
&-\frac{4\lambda\ell}{\pi}[\ci(2\lambda\ell)\cos(2\lambda\ell)+\si(2\lambda\ell)\sin(2\lambda\ell)]\frac{df_p(z)}{dz}\nonumber\\
&+\left.\sum_{n=2}^p\frac{(-2)^n}{\pi^{3/2}}G_{n,n+2}^{4,n-1}\biggl(
\begin{array}{c}
0,\ldots,0,\frac{1}{2},0\\
\frac{1}{2},\frac{1}{2},1,1,1,\ldots,1
\end{array}
\bigg|\lambda^2\ell^2\biggr)\frac{1}{n!}\frac{d^nf_p(z)}{dz^n}\right|_{z=0},
\label{d}
\end{align}
where $G$ is a Meijer function.

Substituting the expansions from Eqs.~(\ref{delta}) and (\ref{exponential}) into the partition function, Eq.~(\ref{Z}), and using the orthogonality relation, Eq.~(\ref{orthogonality}), gives us the \emph{exact} answer to the problem of the closed quantum curve
\begin{equation}
Z={\cal N}\sum_{p_1,p_2,\ldots,p_N}c_{p_1p_2\ldots p_N}d_{p_1}d_{p_2}\cdots d_{p_N}.
\label{answer}
\end{equation}
Thus, unlike the Regge-calculus formulation discussed in the introduction, the closed curve can be treated analytically for arbitrary $N$ without imposing explicit inequalities among the edge lengths.

Just a few terms are necessary to evaluate $Z$, because of the rapid convergence of Eq.~(\ref{answer}) (for $\langle L\rangle$ we need more terms as $\lambda$ grows). Instead of using partial sums to give an approximate expression for $Z$ (or $\langle L\rangle$), we believe that it is more interesting to expand the exact solution, Eq.~(\ref{answer}), for weak and strong cosmological constant. The expansions are calculated in the next section.

\begin{figure}
\centering
\includegraphics[scale=0.3]{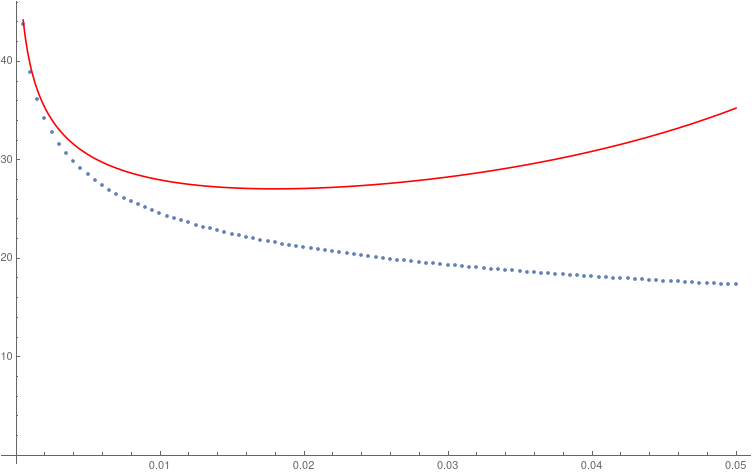}
\includegraphics[scale=0.35]{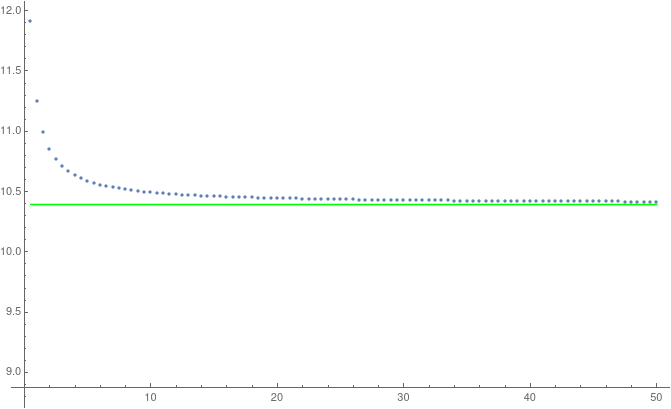}
\caption{Average length $\langle L\rangle/\ell$ for $N=3$ is plotted against $\lambda\ell$ as a dotted curve (in blue) together with the approximations for weak (left figure, Eq.~(\ref{3}), in red) and strong cosmological constant (right figure, Eq.~(\ref{strong}), in green). We split the plot into two panels because the two approximations lie in widely separated ranges of $\lambda\ell$.}
\label{L3}
\end{figure}

\subsubsection{Asymptotic Expansions and Continuum Limit}
For $\lambda\ell\ll 1$ we can approximate the coefficients of Eq.~(\ref{d}) as
\begin{equation}
d_p\approx\delta_{p0}+\frac{4\lambda\ell(-1)^p}{\pi}[\gamma-1+S_p+\ln(2\lambda\ell)],
\label{da}
\end{equation}
where $\gamma$ is the Euler constant, and
\begin{equation}
S_p=\sum_{s=0}^{|p|-1}\frac{1}{2[s/2]+1}
\label{Sp}
\end{equation}
from Appendix~\ref{A}.

For example, with $N=3$, substituting Eq.~(\ref{da}) into Eq.~(\ref{answer}) and keeping terms up to order $\lambda\ell$ we obtain
\[Z\approx 1+\frac{24\lambda\ell}{\pi}[\gamma-1+\ln(4\lambda\ell)],\]
where we have used
\begin{equation}
c_{p00}=
\begin{cases}
1/2\pi,&p=0\\
[(-1)^p-1]/\pi^3p^2,&p\geq 1
\end{cases}
\label{cp00}
\end{equation}
from Eq.~(\ref{c3}), the series
\begin{equation}\label{series}
\sum_{t=0}^\infty\frac{1}{(2t+1)^2}=\frac{\pi^2}{8},
\end{equation}
and the double series
\begin{equation}
\sum_{t=0}^\infty\frac{S_{2t+1}}{(2t+1)^2}=\frac{\pi^2}{4}\ln 2,
\end{equation}
from Appendix~\ref{B}.

Therefore the average length, from Eq.~(\ref{expectation}), is
\begin{equation}
\frac{\langle L\rangle}{\ell}\approx-\frac{24[\gamma+\ln(4\lambda\ell)]}{\pi+24\lambda\ell[\gamma-1+\ln(4\lambda\ell)]}
\label{3}
\end{equation}
which is the curve plotted in red in Fig.~(\ref{L3}). Observe that $\langle L\rangle\to\infty$ in the limit $\lambda\to 0$ (and for $\lambda<0$ too, of course).

For $\lambda\ell\gg 1$ the partition function is dominated by configurations that minimize the total length $L$. So we have to find the minimum of $L$ subject to the constraint of Eq.~(\ref{constraint}). This is easily done using Lagrange multipliers and the answer is trivial: all angles are equal, thus $\phi_n=2\pi/N$ and
\begin{equation}
\frac{\langle L\rangle_\text{min}}{\ell}=2N\tan\left(\frac{\pi}{N}\right)
\label{strong}
\end{equation}
is the minimum length.

The case of the triangle, $N=3$, is plotted in Fig.~(\ref{L3}) as a constant curve in green, where $\langle L\rangle_\text{min}=6\sqrt{3}\ell\approx 10.39\ell$ is the perimeter of an equilateral triangle with an inscribed circle of radius $\ell$. Note that the closed curve is different from an open one \cite{Hamber}, where the open curve could shrink to zero length. Here, the constraint of Eq.~(\ref{constraint}) together with the assumption $\ell=$ a constant, fixes a minimum length given by Eq.~(\ref{strong}).

The curves get steeper as the number of edges $N$ grows in the region of small $\lambda$. From Eq.~(\ref{strong}) we have $\langle L\rangle\to 2\pi\ell$ as $N\to\infty$ in the region of large $\lambda$. Therefore, in the continuum limit, our model behaves like an ``infinite step''
\begin{equation}
\frac{\langle L\rangle}{\ell}=
\begin{cases}
\infty,&\lambda\leq 0\\
2\pi,&\lambda> 0
\end{cases}
\label{continuum}
\end{equation}
in the closed curve case for pure gravity.

The fixed radius therefore changes the continuum limit: for positive cosmological constant, the closed geometry approaches the circumference $2\pi\ell$ rather than diverging as in the Regge calculus. In the presence of a scalar field this result will change slightly.

\subsection{Scalar Field}
Now, let us put a scalar field $-\infty<\varphi_n<\infty$, for $n=1,2,3,\dots,N$, at each tangency point of Fig.~(\ref{curve}). In this case, the action for the scalar field is \cite{Christ}
\[I=\sum_{n=1}^N\left[\frac{(\varphi_{n+1}-\varphi_n)^2}{2l_n}+\frac{m^2(l_n+l_{n-1})\varphi_n^2}{4}\right]\]
and so we have Gaussian integrals. Therefore, the integration over $\varphi$ can be performed exactly for a general $N$ and arbitrary mass $m$.

But to calculate analytically the following integrations over the angles $\phi_n$, let us assume that the mass $m$ is small and the measure is
$\prod_{n=1}^Nd\varphi_n/\sqrt{l_n}$, see Ref.~\cite{Unz}. To first order in $m\ell$, after performing the Gaussian integrals, Eq.~(\ref{Z}) is replaced by
\[Z\approx{\cal N}\left(\prod_{n=1}^N\int_0^\pi\frac{d\phi_n}{2\pi}\right)\frac{\delta\left(\sum_{n=1}^N\phi_n-2\pi\right)}{\sum_{n=1}^N\tan\frac{\phi_n}{2}}\exp\left(-2\lambda\ell\sum_{n=1}^N\tan\frac{\phi_n}{2}\right)\]
and the coefficients in Eq.~(\ref{c}) can be replaced by
\[c_{p_1p_2\ldots p_N}=\int_{-\pi}^\pi\frac{d\phi_1}{2\pi}\int_{-\pi}^\pi\frac{d\phi_2}{2\pi}\ldots\int_{-\pi}^\pi\frac{d\phi_N}{2\pi}\frac{\delta\left(\sum_{n=1}^N|\phi_n|-2\pi\right)}{\sum_{n=1}^N\tan\frac{|\phi_n|}{2}}\chi_{p_1}^*(\phi_1)\chi_{p_2}^*(\phi_2)\ldots\chi_{p_N}^*(\phi_N).\]
\subsubsection{Asymptotic Expansions and Continuum Limit}
\begin{figure}
\centering
\includegraphics[scale=0.64]{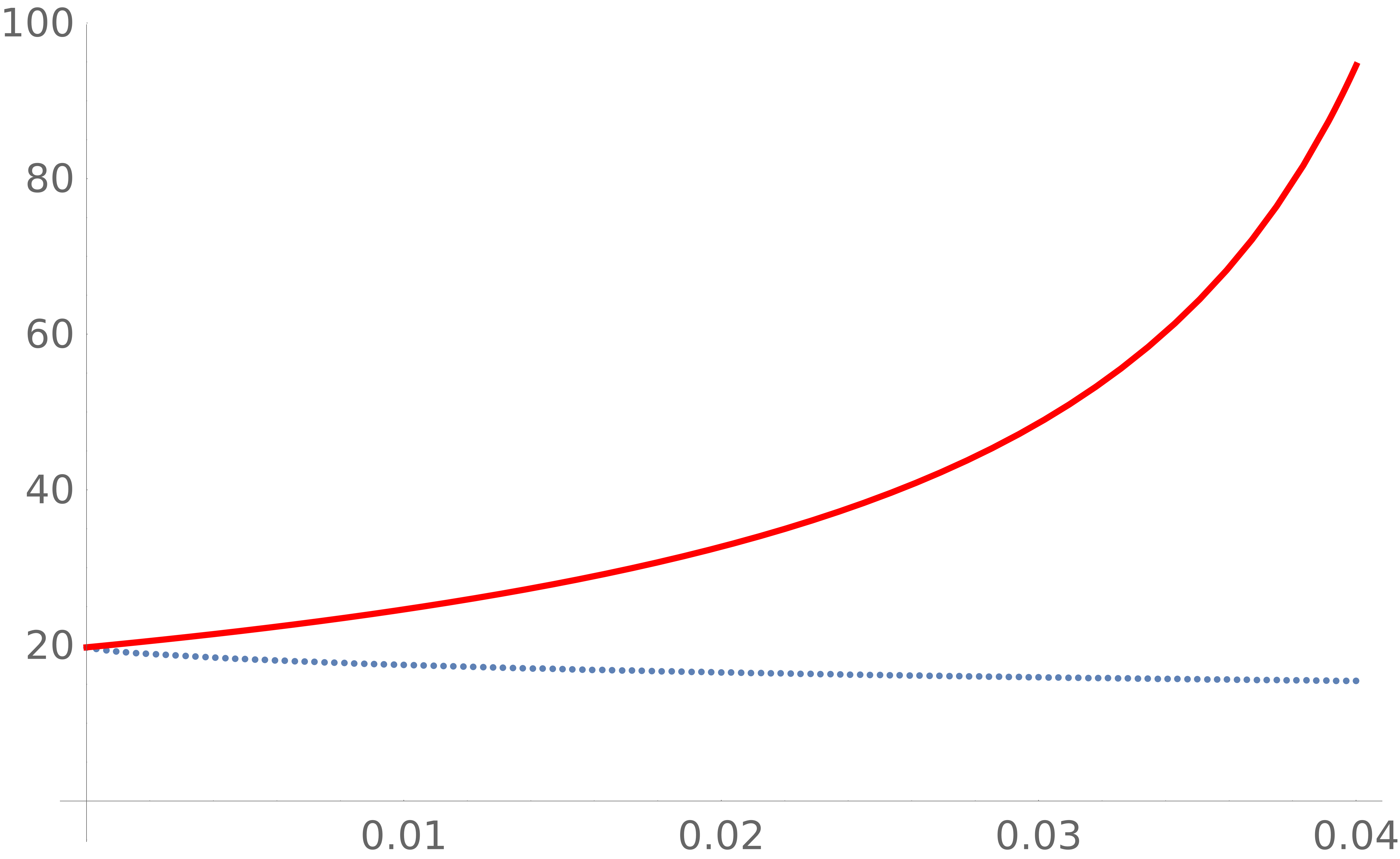}
\includegraphics[scale=0.65]{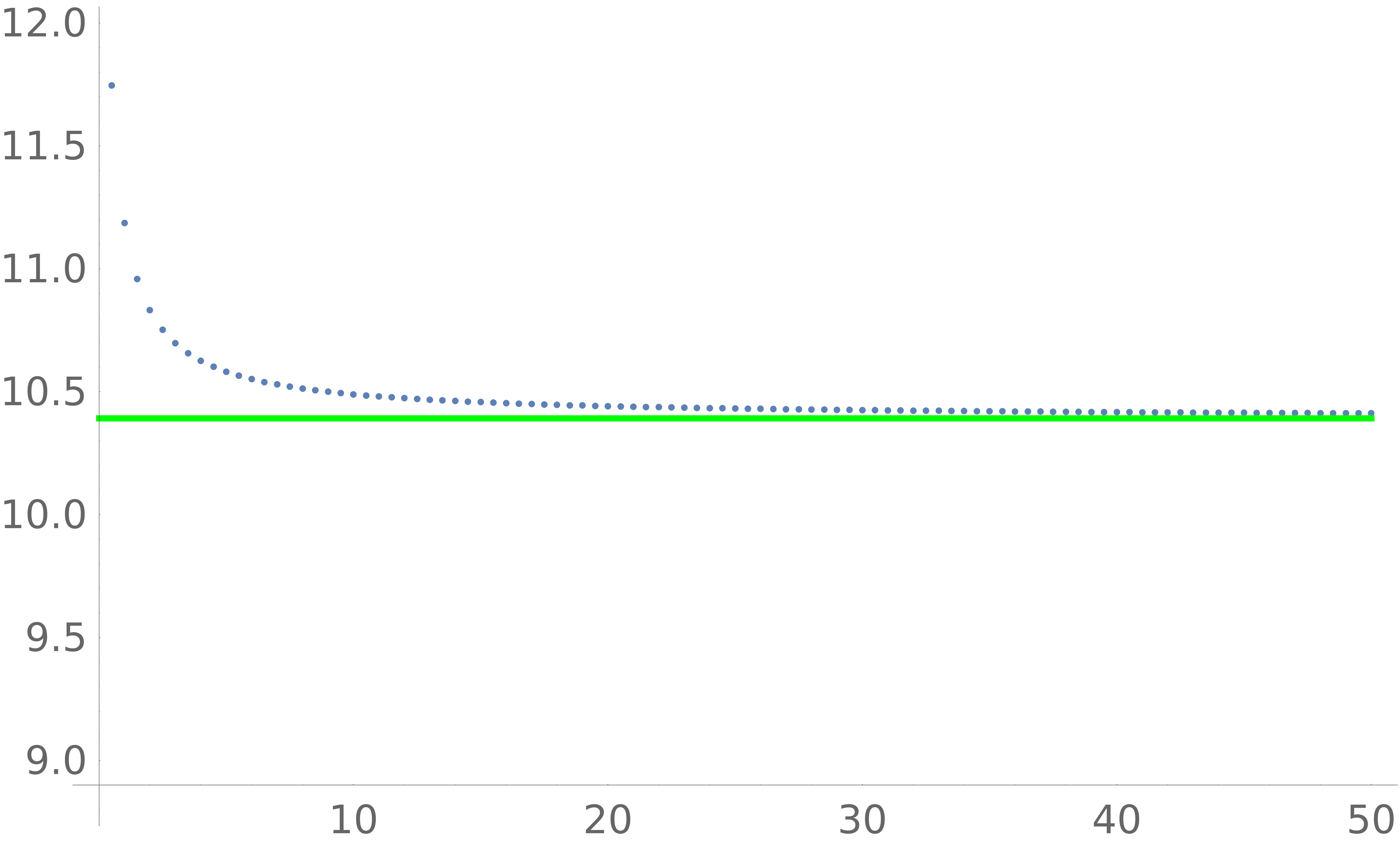}
\caption{Same figure as Fig.~(\ref{L3}) but now with a scalar field.}
\label{L3n}
\end{figure}

For simplicity, let us take $N=3$ again. Since we are interested in the weak $\lambda$ regime (the strong limit does not change in the presence of a scalar field, see Fig.~(\ref{L3n})), it suffices to calculate
\[c_{p00}=
\begin{cases}
2(3\ln 2-2)/\pi^2,& p=0\\
[1/|p|-4+(8|p|+2)\Phi(-1,1,|p|+1)](-1)^p/\pi^2,& p\geq 1
\end{cases}\]
where $\Phi$ is a Lerch transcendent.

Therefore
\[Z\approx 1-\frac{\pi\lambda\ell}{2(3\ln 2-2)},\]
where we have used
\[\sum_{p=1}^\infty\left[\frac{1}{p}-4+(8p+2)\Phi(-1,1,p+1)\right]=-(3\ln 2-2)\]
and
\[\sum_{p=1}^\infty\left[\frac{1}{p}-4+(8p+2)\Phi(-1,1,p+1)\right]S_p=-\frac{\pi^2}{24}\]
from a calculation analogous to that in Appendix~\ref{B}, but involving considerably more algebra.

So the average length, from Eq.~(\ref{expectation}), is
\begin{equation}
\frac{\langle L\rangle}{\ell}\approx\frac{\pi}{2(3\ln 2-2)-\pi\lambda\ell}
\end{equation}
which is the curve plotted in red in Fig.~(\ref{L3n}). Observe that a small scalar mass changes the limit $\lambda\to 0^+$ to $\pi/2(3\ln 2-2)\approx 19.77$, a finite value.

Numerical calculations using the exact partition function, that is, without imposing that the scalar field mass $m$ is small, show that as the value of the mass $m$ increases, the limit $\lambda\to 0^+$ becomes closer to Eq.~(\ref{strong}).

Therefore, in the presence of a scalar field, regardless of the mass $m$, the continuum limit $N\to\infty$ is now
\begin{equation}
\frac{\langle L\rangle}{\ell}=
\begin{cases}
\infty,&\lambda< 0\\
2\pi,&\lambda\geq 0
\end{cases}
\label{continuum_new}
\end{equation}
which we can compare to Eq.~(\ref{continuum}). Notice that the difference is subtle: the only change occurs at $\lambda=0$.

\section{Conclusions}
A path integral, as in Eqs.~(\ref{open}) or (\ref{Z}), is a ``sum over paths'', that in the case of quantum gravity means an integration in the space of all possible metrics $g$. In a 1D lattice, with edge lengths $l_n$, since $g$ has only one component, $g=(g_{11})$, we can choose $g_{11}(l_n)$ to be the same as the length squared, thus, $g_{11}(l_n)=l_n^2$. Therefore, it is natural to consider integrals of $l_n$. But what integration measure is the right one?

As far as we know, there is no general consensus about the measure, $Dg=\prod_ndl_nf(l_n)$, where $f(l_n)$ is a function of only $l_n$ for a local measure\footnote{For a non-local measure, $f$ is usually chosen as a function of the first neighbors. In 1D, this is $f(l_{n-1},l_n,l_{n+1})$.}. For example, in Regge calculus \cite{Hamber} they choose $f(l_n)=2l_n^{\sigma+1}$ for some number $\sigma$. Besides the ambiguity in the value of $\sigma$, there are more difficulties with Regge calculus such as the problem with inequalities, ambiguity in the value of the short distance cutoff $\epsilon$, numerical precision, etc.

The main idea of our simple model is to work with the angles, $\phi_n$, and use the unique normalized Haar measure $\prod_n d\phi_n/2\pi$ for U(1), which is a compact group. The open curve was easy to calculate, see Eq.~(\ref{open}). But in the slightly more realistic case of the closed curve, Eq.~(\ref{Z}), there were some problems.

The first problem is that we cannot use perturbation theory with an exponential $\exp(-\lambda L)=1-\lambda L+\lambda^2 L^2/2+\ldots$ that diverges $\int Dg\,L=\int Dg\,L^2=\ldots=\infty$ at each order in $\lambda$, see Ref.~\cite{Iwata}. This is similar to the quantum field theory expansions for quantum gravity, see for example Ref.~\cite{Paszko}, where we have to absorb a new divergence at each order in Newton's constant, which is impossible and makes these theories non-renormalizable.

Because we are using the Haar measure, an alternative is to expand the Dirac delta function and the exponentials in characters, $\chi_p(\phi_n)$, which brings us to another problem: the \emph{global} constraint, Eq.~(\ref{constraint}), involves \emph{all} the angles $\phi_n$ and thus cannot be treated with the usual delta function expansion $\delta(\phi)=\sum_p\chi_p(\phi)/2\pi$, see footnote \ref{Dirac}.

To solve this second problem, we used a new expansion, Eq.~(\ref{delta}), that multiplies all characters, one for each angle $\phi_n$. The result in the closed case is very interesting: the average length, $\langle L\rangle$, varies from a minimum value in the limit $\lambda\to\infty$, Eq.~(\ref{strong}), to infinity in the limit $\lambda\to 0$, Eq.~(\ref{3}), see Fig.~(\ref{L3}).

The continuum limit is also different for the closed curve, as we can see in Eq.~(\ref{continuum}). We have a nontrivial limit, that is, an infinite step at $\lambda=0$. Of course, this is a consequence of the circle inside the curve in our model, see Fig.~(\ref{curve}). This circle is not present in Regge calculus \cite{Hamber} and so it has a trivial continuum limit in the closed case as well, $\langle L\rangle\to\infty$, for any value of $\lambda$.

In summary, in the case of an open curve, the size of the universe is infinite, regardless of the value of $\lambda$. This seems unrealistic, since $\lambda$ controls the expansion of the universe in more realistic models. But in the closed case, we have a universe with finite size for $\lambda>0$.

Notice that the radius $\ell$ of the U(1) circle is related to the minimum average length $\langle L\rangle_\text{min}$ of Eq.~(\ref{strong}). This minimum length corresponds to a shape that has the highest degree of symmetry, which helps us understand how similar scenarios might work in higher dimensions.

For example, in 2D, we can use a two-sphere with a constant radius $\ell$ within a surface made of triangles. As the value of $\lambda$ increases, this setup approaches a regular surface. Similarly, in 3D, a three-sphere can be surrounded by tetrahedra, and so on\textellipsis

It is important to note that adding a scalar field does not significantly change the results, as shown in Eq.~(\ref{continuum_new}), regardless of its mass. It is natural that mass is not coupled to gravity in 1D or even in 2D.

The constant $\ell$ might be related to the Planck length $\ell_p$, which is connected to Newton's constant. However, we still need to confirm whether such a relationship exists. It seems likely that in 3D or more, we can explore this further, as curvature becomes significant and the Newtonian potential is non-zero.

Currently, the 2D extension of the present model, where the corresponding construction is based on a tetrahedron with an inscribed sphere of fixed radius $\ell$, is being developed in Ref.~\cite{Paszko2}.
\appendix
\section{Asymptotic approximation}\label{A}
For small $x=2\lambda\ell$ the coefficients in Eq.~(\ref{d}) can be approximated by
\[d_p\approx\delta_{p0}+\frac{2x(-1)^p}{\pi}(\gamma-1+\ln x+S_p),\]
where
\[S_0=0,\;S_1=1,\;S_2=2,\;S_3=\frac{7}{3},\;S_4=\frac{8}{3},\;S_5=\frac{43}{15},\;S_6=\frac{46}{15},\;S_7=\frac{337}{105},\;S_8=\frac{352}{105},\;\ldots\]

This sequence can be written as
\begin{align*}
1&=1\\
2&=1+1\\
\frac{7}{3}&=1+1+\frac{1}{3}\\
\frac{8}{3}&=1+1+\frac{1}{3}+\frac{1}{3}\\
\frac{43}{15}&=1+1+\frac{1}{3}+\frac{1}{3}+\frac{1}{5}\\
\frac{46}{15}&=1+1+\frac{1}{3}+\frac{1}{3}+\frac{1}{5}+\frac{1}{5}\\
\frac{337}{105}&=1+1+\frac{1}{3}+\frac{1}{3}+\frac{1}{5}+\frac{1}{5}+\frac{1}{7}\\
\frac{352}{105}&=1+1+\frac{1}{3}+\frac{1}{3}+\frac{1}{5}+\frac{1}{5}+\frac{1}{7}+\frac{1}{7}\\
&\vdots
\end{align*}
which is equal to
\[S_p=\sum_{s=0}^{p-1}\frac{1}{2[s/2]+1},\]
where $[s/2]$ is the integer part of $s/2$.
\section{Double series}\label{B}
To find the sum of $S_p/p^2$, see Eq.~(\ref{Sp}), for $p=2t+1$ an odd integer, we note that
\begin{gather*}
\sum_{t=0}^\infty\frac{S_{2t+1}}{(2t+1)^2}=\sum_{t=0}^\infty\frac{1}{(2t+1)^2}\sum_{s=0}^{2t}\frac{1}{2[s/2]+1}\\
=2\left[1+\frac{1}{3^2}\left(1+\frac{1}{3}\right)+\frac{1}{5^2}\left(1+\frac{1}{3}+\frac{1}{5}\right)+\frac{1}{7^2}\left(1+\frac{1}{3}+\frac{1}{5}+\frac{1}{7}\right)+\ldots\right]-\left(1+\frac{1}{3^3}+\frac{1}{5^3}+\frac{1}{7^3}+\ldots\right)\\
=2\sum_{m=0}^\infty\frac{1}{(2m+1)^2}\sum_{n=0}^m\frac{1}{2n+1}-\sum_{m=0}^\infty\frac{1}{(2m+1)^3}\\
=2\sum_{m=0}^\infty\frac{1}{(2m+1)^2}\left[\ln 2+\frac{\gamma}{2}+\frac{\psi(m+3/2)}{2}\right]-\frac{7}{8}\zeta(3),
\end{gather*}
where $\psi$ is the digamma function and $\zeta$ is the Riemann zeta function.

Using Eq.~(\ref{series}) and Eq.~(55.2.10) from the addendum of Ref.~\cite{Hansen}
\begin{gather*}
\sum_{m=0}^\infty\frac{\psi(m+3/2)-\psi(1/2)}{(2m+1)^{2n}}=(1-2^{-2n-1})\zeta(2n+1)+2(1-2^{-2n})\zeta(2n)\ln 2\\
-2^{-2n}\sum_{k=1}^{n-1}(2^{2k}-1)\zeta(2k)\zeta(2n-2k+1)
\end{gather*}
valid for $n=1,2,3,\ldots$ (for $n=1$, omit the sum in the right-hand side), we obtain the double series
\[\sum_{t=0}^\infty\frac{S_{2t+1}}{(2t+1)^2}=\frac{\pi^2}{4}\ln 2.\]

\end{document}